\documentclass[apj]{emulateapj}
\usepackage{amsmath}
\usepackage{graphicx}
\usepackage{comment}
\newcommand{\panstarrs}[0]{Pan-{\sc starrs}}
\newcommand{\panstarrss}[0]{Pan-{\sc starrs} }
\newcommand{\besancon}[0]{Besan{\c c}on }
\newcommand{\citepy}[1]{\citeauthor{#1} (\citeyear{#1})}
\begin{document}
\slugcomment{accepted for publication into PASP Vol. 122 No. 897}
\shorttitle{Building an Optimal Census of the Solar Neighborhood with Pan-STARRS Data}
\shortauthors{Beaumont \& Magnier}
\title{Building an Optimal Census of the Solar Neighborhood \\ with Pan-STARRS Data}
\author{Christopher N. Beaumont \& Eugene A. Magnier}
\email{beaumont@ifa.hawaii.edu, eugene@ifa.hawaii.edu}
\affil{Institute for Astronomy, University of Hawai'i at Manoa \\ 2680 Woodlawn Drive, Honolulu HI 96822}
\begin{abstract}
We estimate the fidelity of solar neighborhood ($D < 100\,$pc) catalogs soon to be derived from \panstarrss astrometric data. We explore two quantities used to measure catalog quality: completeness, the fraction of desired sources included in a catalog; and reliability, the fraction of entries corresponding to desired sources. We show that the main challenge in identifying nearby objects with \panstarrss will be reliably distinguishing these objects from distant stars, which are vastly more numerous. We explore how joint cuts on proper motion and parallax will impact catalog reliability and completeness. Using synthesized astrometry catalogs, we derive optimum parallax and proper motion cuts to build a census of the solar neighborhood with the \panstarrss $3 \pi$ Survey. Depending on the Galactic latitude, a parallax cut $\pi/\sigma_\pi >5$  combined with a proper motion cut ranging from $\mu/\sigma_\mu >$1--8 achieves 99\% reliability and 60\% completeness.
\end{abstract}
\keywords{Data Analysis and Techniques}
\maketitle

\section{Introduction}
The next generation of synoptic sky surveys (\panstarrs, \citealt{Kaiser04}; {\sc lsst}, \citealt{Tyson02}) will have the necessary astrometric precision and time cadence to measure parallax for sources in the solar neighborhood ($D \lesssim 100$\,pc). These parallax measurements will be among the most useful signals in building a census of the local stellar environment. Other popular methods for identifying nearby objects (e.g., selection based on proper motion or color-magnitude constraints) do not unambiguously constrain distance, and require follow-up spectroscopic and/or astrometric observations. Furthermore, such selection criteria are ineffective at discovering sources with extreme properties. For example, a bright L dwarf within 10 pc was only recently discovered in the Sloan Digital Sky Survey; it was not found in previous color-based searches because of its unusually blue spectral energy distribution (SED) \citep{Schmidt10, Bowler10}. Deep, all-sky parallax surveys promise to identify nearby objects, regardless of any unusual photometric properties, without follow-up telescope investment.

Currently, the Hipparcos and Tycho catalogs are the only all-sky parallax databases \citep{Perryman97}. While these catalogs are precise, they are also shallow ($m_{\rm v} \lesssim 12$) and hence biased towards bright, high mass objects. The \panstarrss $3 \pi$ Survey \citep{Magnier08} is a deep, ground-based synoptic survey that will significantly extend the Hipparcos catalogs. The $3 \pi$ Survey will observe 3/4 of the sky 60 times over the course of 3 years, and will provide 5-band (grizy) photometry for objects in the range $15.3 < m_r < 22.8$\footnote{Expected detection limits in the gizy filters are 23.4, 22.2, 21.6, and 20.1, respectively}. The survey has an astrometric error floor of 10 milliarcseconds (mas) per exposure, resulting in a stacked parallax sensitivity of $\sigma_\pi \sim 2$\,mas per source.

This survey will yield a much more complete census of the low mass objects in the solar neighborhood. The first \panstarrss telescope (PS1\footnote{Four telescopes are planned for the long-term \panstarrss mission}) saw first light in 2007, and the PS1 science mission began on May 1, 2010.

The $3 \pi$ Survey will be deeper but less astrometrically precise than the Hipparcos catalogs. This poses a unique challenge for identifying nearby objects in \panstarrss data. Distant objects are far more numerous than nearby ones and, because of astrometric measurement errors, will occasionally display false parallax signals. Thus interlopers -- distant objects with spuriously large parallax measurements -- will outnumber accurate but small parallax measurements. The focus of this paper is to quantify this effect, and explore how additional proper motion cuts can break the ambiguity between interlopers and legitimate solar neighbors.

Interloper confusion is a manifestation of the Lutz-Kelker bias \citep{Lutz73}, so we begin with a review of that phenomenon in \S\ref{sec:lk}. A preliminary analysis of the Lutz-Kelker bias suggests that sources more distant than 80\,pc will be difficult to reliably extract from \panstarrss data using parallax alone; this is a stricter limit than the 100\,pc horizon assumed in previous estimates of \panstarrs' solar neighborhood performance (\citealt{Magnier08}; \citealt{Dupuy09}). Thus, in \S\ref{sec:pm} we simulate to what extent additional proper-motion cuts can suppress interlopers, and what effect these selection criteria have on catalog completeness. 

We find that cuts based on proper motion and parallax are more effective than cuts based on parallax alone. We estimate that high reliability (99\%) catalogs derived from \panstarrss data will be approximately 60\% complete, and that catalogs with 50\% reliability will be 85\% complete.

\section{The Lutz-Kelker Bias}
\label{sec:lk}

\begin{figure}
\includegraphics[width=3.5in]{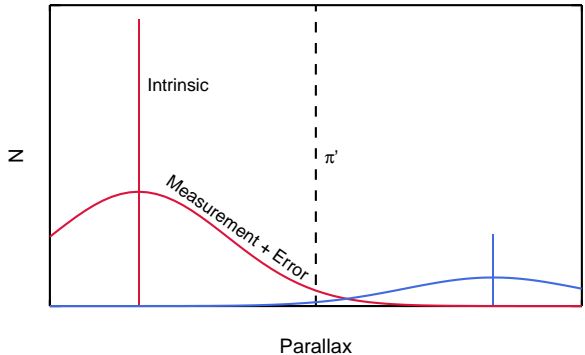}
\caption{A schematic illustration of the Lutz-Kelker bias. The bars show a hypothetical parallax histogram for two classes of objects, the more distant of which is also more numerous. When the parallax of each source is observed in the presence of noise, the resulting histograms spread out into the gaussian curves shown in the figure. The group of objects measured to have parallaxes of $\pi'$ (the dotted line) consists mostly of the more-numerous (red) sources. Thus, on average, $\pi'$ is an overestimate for sources measured to have this parallax.}
\label{fig:cartoon}
\end{figure}

\citepy{Trumpler53} first noted that parallax measurements tend to systematically underestimate true distances. This effect was subsequently analyzed in detail by \citepy{Lutz73}, and has become known as the Lutz-Kelker bias (hereafter LK). The LK bias is a consequence of the fact that distant stars are more numerous than nearby ones. To understand this on an intuitive level, consider the illustration in Figure \ref{fig:cartoon}. The red and blue bars denote parallax histograms for two classes of objects: a large group of distant objects (red bar), and a smaller group of nearby objects (blue bar). The curves show how these histograms change when the parallax is measured in the presence of noise -- objects scatter away from their true values. Now consider the parallax value $\pi'$ located midway between each bar (the dotted line). There are more objects from the red group at this position than from the blue group. Thus, objects with measured parallaxes of $\pi'$ have, on average, actual parallaxes smaller than $\pi'$. This illustration can immediately be generalized to continuous distributions of stellar distances; when the distribution of stars increases with distance, measured parallaxes underestimate true distances.

To quantify this effect, we follow Haywood's (\citeyear{Smith87}) approach and use Bayes' theorem to express the probability distribution of an object's distance, constrained both by a parallax measurement and a-priori knowledge of how objects are distributed in space:
\begin{equation}
\label{eq:Bayes}
P(D | \pi) \propto P(\pi | D) P(D) = \mathcal{N}(\pi; D^{-1}, \sigma_\pi) P(D)
\end{equation}
$P(D | \pi)$ is the posterior probability, or the probability distribution of an object's true distance, constrained by a parallax measurement of $\pi$. $P(\pi | D)$ is the likelihood of measuring $\pi$, assuming the true distance is D. The likelihood is derived from measurement errors, which we assume to be Gaussian. In this case, the likelihood is given by 
$\mathcal{N}(\pi; D^{-1}, \sigma_\pi)$, the Normal Distribution with mean $D^{-1}$ and standard deviation $\sigma_\pi$, evaluated at $\pi$. The prior distribution $P(D)$ is the intrinsic distribution of objects as a function of distance, or $dN/dD$. In an observation of solid angle $\Omega$, this number is given (to within a normalization constant) by
\begin{equation}
\label{eq:prior}
dN = \Omega D^2 n(D) dD
\end{equation}
Here, $n(D)$ is the number density of objects as a function of $D$. 

\begin{figure}
\includegraphics[width=3.5in]{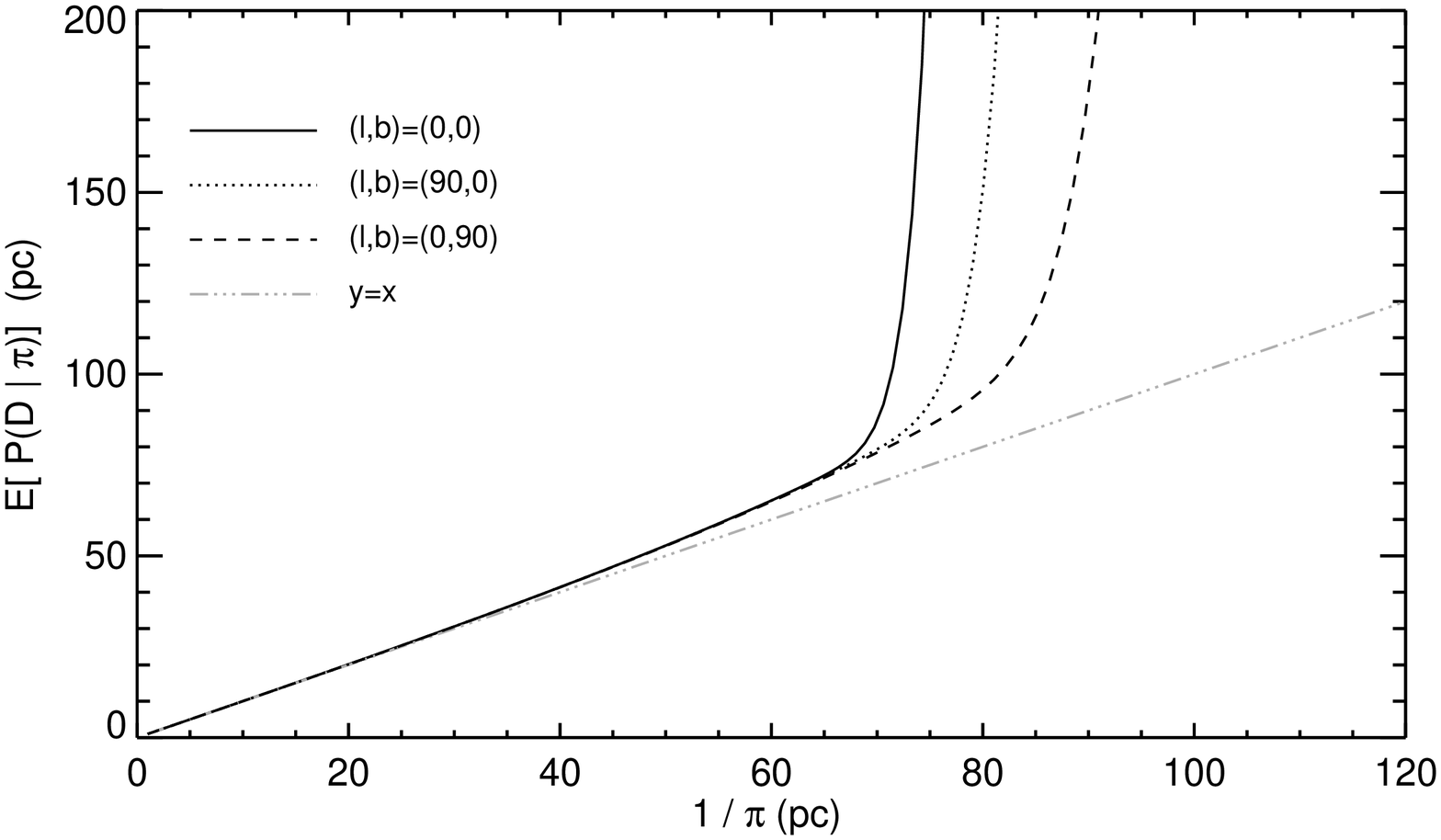}
\includegraphics[width=3.5in]{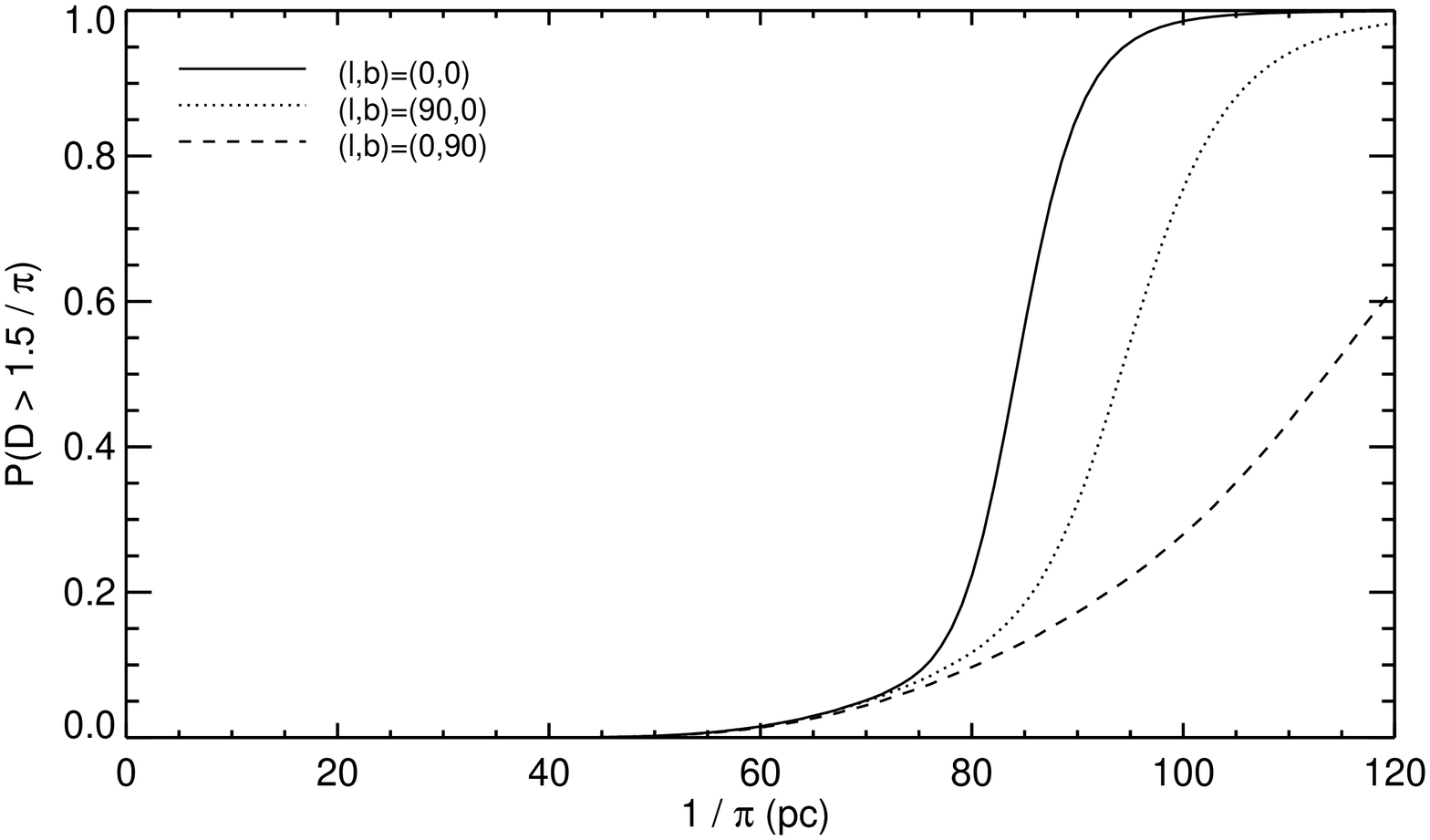}
\caption{The behavior of the Bayesian estimator given by Equation \ref{eq:Bayes}, assuming a measurement error of $\sigma_\pi = 2$\,mas. The three curves trace different lines of sight; through the Galactic center (solid line), in the Galactic plane along $\ell = 90^o$, and out of the Galactic plane. Top: the mean distance of objects measured to have a parallax of $1/\pi$; this diverges from the value of $1/\pi$ as that number grows. Bottom: the probability that an object's distance is \textgreater 150\% the value implied by the parallax measurement.}
\label{fig:lk}
\end{figure}

Figure \ref{fig:lk} shows the behavior of Equation \ref{eq:Bayes}, assuming that $\sigma_\pi$ = 2 milliarcseconds (mas); this is the estimated maximum precision of the \panstarrss 3$\pi$ Survey (see the Appendix). The top panel plots the mean value of Equation \ref{eq:Bayes} for three different expressions for $n(D)$. They correspond to three lines of sight: one towards the Galactic center, one out of the Galactic plane, and one away from the Galactic center. Equations for the stellar number density are derived from the Galactic parameters given in \citepy{Bochanski09}. These three lines of sight bracket the behavior of $n(D)$, which increases most rapidly towards the Galactic center, and decreases most rapidly out of the plane. Note that the mean distance of Equation \ref{eq:Bayes} diverges from the parallax estimate when $1/\pi \gtrsim 70$\, pc.

The bottom panel plots a similar function: the probability that the true distance is more than 150\% the value implied by the parallax. Again, this probability becomes substantial around $1/\pi \gtrsim 80$\,pc. Previous estimates of \panstarrss yield \citep{Magnier08, Dupuy09} have assumed that parallax catalogs will be complete out to 100\,pc: a volume twice as large as the 80\,pc sphere at which interloper confusion becomes problematic.

The limitations of parallax-selected catalogs at moderate distances raises the question of whether other quantities might help to increase catalog reliability, without excessively compromising completeness. Two promising quantities include source color (since the solar neighborhood is dominated by red, low mass stars) and proper motion. In the present work, we estimate the quality of solar neighborhood catalogs derived from the \panstarrss $3 \pi$ Survey, when both parallax and proper motion are used to identify nearby objects. 

\section{Proper Motion Cuts}
\label{sec:pm}

The Bayesian formalism above can, in principle, be extended to study how proper motions further constrain distance. However, doing so requires defining a prior function that realistically models the joint distribution of stellar distances and proper motions (that is, $P(D)$ is replaced by $P(D, \mu)$ in Equation \ref{eq:Bayes}). Furthermore, we wish to employ a realistic astrometric error model, which depends on source flux. In total, a fully analytic Bayesian treatment would require defining a 8-dimensional prior distribution (distance, proper motion in two dimensions, and flux in 5 \panstarrss filters), and marginalizing over 7 of them. We opt instead of a more straightforward numerical approach using simulated \panstarrss source catalogs. This is more tractable, and allows us to utilize existing research into synthesized models of the Galaxy.

Our approach is as follows. First, we generate catalogs of intrinsic source properties (luminosity, color, velocity, and distance). We then assign astrometric uncertainties and calculate, given a certain set of parallax and proper motion selection criteria, the probability that each source would be measured to satisfy these cuts. By summing these probabilities, we arrive at the expected number of sources included in the solar neighborhood catalog. Finally, we can compare this number to the actual number of nearby objects in the catalogs to measure the performance of the classification strategy.

Our synthesized catalogs have been generated from the \besancon model of the Galaxy \citep{Robin03}\footnote{http://model.obs-besancon.fr/}. This model accurately represents our current knowledge of how stars are distributed in space, velocity, mass, and age. Table \ref{tab:besancon} summarizes the parameters we have used when querying the \besancon model service. We have generated catalogs for three different positions on the sky: (l,b) = (90, 0), (180, 0), and (0, 90). We avoid analyzing lines of sight near Galactic center for three reasons: the very high density of sources makes the analysis cumbersome; the \panstarrss survey will be confusion-limited towards this region and hence difficult to analyze statistically; and the region is compact on the sky, and hence not a significant source of nearby objects.

\begin{deluxetable}{ll}
\tablecolumns{2}
\tablewidth{0in}
\tabletypesize{\scriptsize}
\tablecaption{Parameters of the \besancon model queries \label{tab:besancon}}
\tablehead{\colhead{Parameter} & \colhead{Value}}
\startdata
Filter System & CFHT-Megacam \\
Distance interval & $0 < D < 50$\,kpc \\
Small / Large field & Small \\
Positions ($\ell,b$) & (90,0), (180,0), (0,90) \\
Solid angle\tablenotemark{a} & 4, 20, 1000 deg$^2$ \\
Extinction Law & 0 mag kpc$^{-1}$ \\
Apparent magnitude filter\tablenotemark{b} &r \textless ~26 \\
Noise & none \\
\enddata
\tablenotetext{a}{Solid angles were chosen to generate large but manageable catalogs, containing 4-6 million entries.}
\tablenotetext{b}{This range is wide enough to include all stars detected in the \panstarrss filter system.}
\end{deluxetable}

For each entry in these catalogs, we calculate the (noise-free) parallax, proper motion, and magnitudes in the \panstarrss (g,r,i,z,y) filter system. We also determine the estimated error on each of these quantities. The details of this process are covered in the Appendix. 

We next calculate the probability that any object would be selected by some cut on parallax and proper motion, when those sources are measured in the presence of realistic noise. We assume that all measurement errors are Gaussian, so that the probability of measuring a parallax greater than some threshold $\pi_t$ is obtained directly from the Gaussian cumulative distribution function. We also consider the probability that the proper motion (that is, $\mu \equiv \sqrt{\mu_x^2 + \mu_y^2}$) exceeds some threshold $\mu_t$. When $\mu_x$ and $\mu_y$ are measured with uncertainty $\sigma_\mu$, this probability is given by
\begin{equation}
P(\mu > \mu_t) = \int_{\mu_t^2/\sigma_\mu^2}^{\infty} \chi^2\bigg({\rm x},~ 2, ~\frac{\mu^2}{\sigma_\mu^2}\bigg)~d{\rm x}
\end{equation}
where $\chi^2({\rm x}, \nu, \lambda)$ is the non-central chi-square distribution, with $\nu$ degrees of freedom and non-centrality parameter $\lambda$, evaluated at x \citep{Patnaik49}. 

\subsection{Defining Completeness and Reliability}
The task at hand is to choose some combination of selection criteria on $\pi, \mu$, and the errors on these quantities, to optimize the solar neighborhood catalog. To do so, we define the quantities reliability and completeness to describe catalog fidelity. These are application-dependent concepts, and different definitions lead to different ``optimal'' selection criteria. We wish is to assess the utility of \panstarrss data in identifying objects within the solar neighborhood, so we use the following operational definitions for completeness and reliability:

The completeness of the \panstarrss solar neighborhood catalog is defined to be the fraction of observable objects within 100\,pc that are included in the catalog. An object is said to be observable if it is detected but not saturated in at least one filter at each epoch. The value of 100\,pc is chosen for consistency with previous studies \citep{Dupuy09}, and represents a $\sim 5 \sigma$ parallax detection for bright sources (see the Appendix). The reliability is defined as the fraction of the catalog occupied by sources with true distances less than 200\,pc. Note the ``buffer'' from 100-200\,pc; this is designed so that only sources substantially more distant than 100\,pc degrade the reliability. Including a source within 200\,pc does not compromise a catalog's census of nearby objects. However, a source beyond 200\,pc displays a parallax signal measured at $<2.5 \sigma$ significance, and hence would not be considered a ``detection'' by most criteria.

In summary, the completeness gives the probability that a nearby source is in the catalog (provided it was detected photometrically); the reliability gives the probability that a random catalog object is indeed in the solar neighborhood. 

We have subjected the catalogs described in the previous section to several filtering strategies, and measured the completeness and reliability for each strategy.  Each strategy combines three different cuts: an absolute parallax cut ($1/\pi < 200$\,pc, or $\pi > 5$\, mas), and a cut in relative parallax and proper motion ($\pi/\sigma_\pi > x$, $\mu/\sigma_\mu > y$). We sampled a grid of strategies, ranging from $0 < x < 7$ in steps of 1 and $0 < y < 30$ in steps of 5. We also sampled a sub-region $2.5 < x < 5$ in steps of 0.2 and $0 < y < 9$ in steps of 0.5. In each simulation, any object meeting all criteria (absolute parallax, relative parallax, and relative proper motion) is labeled as a nearby source.

\section{Results}

\begin{figure}
\includegraphics[height=2.8in]{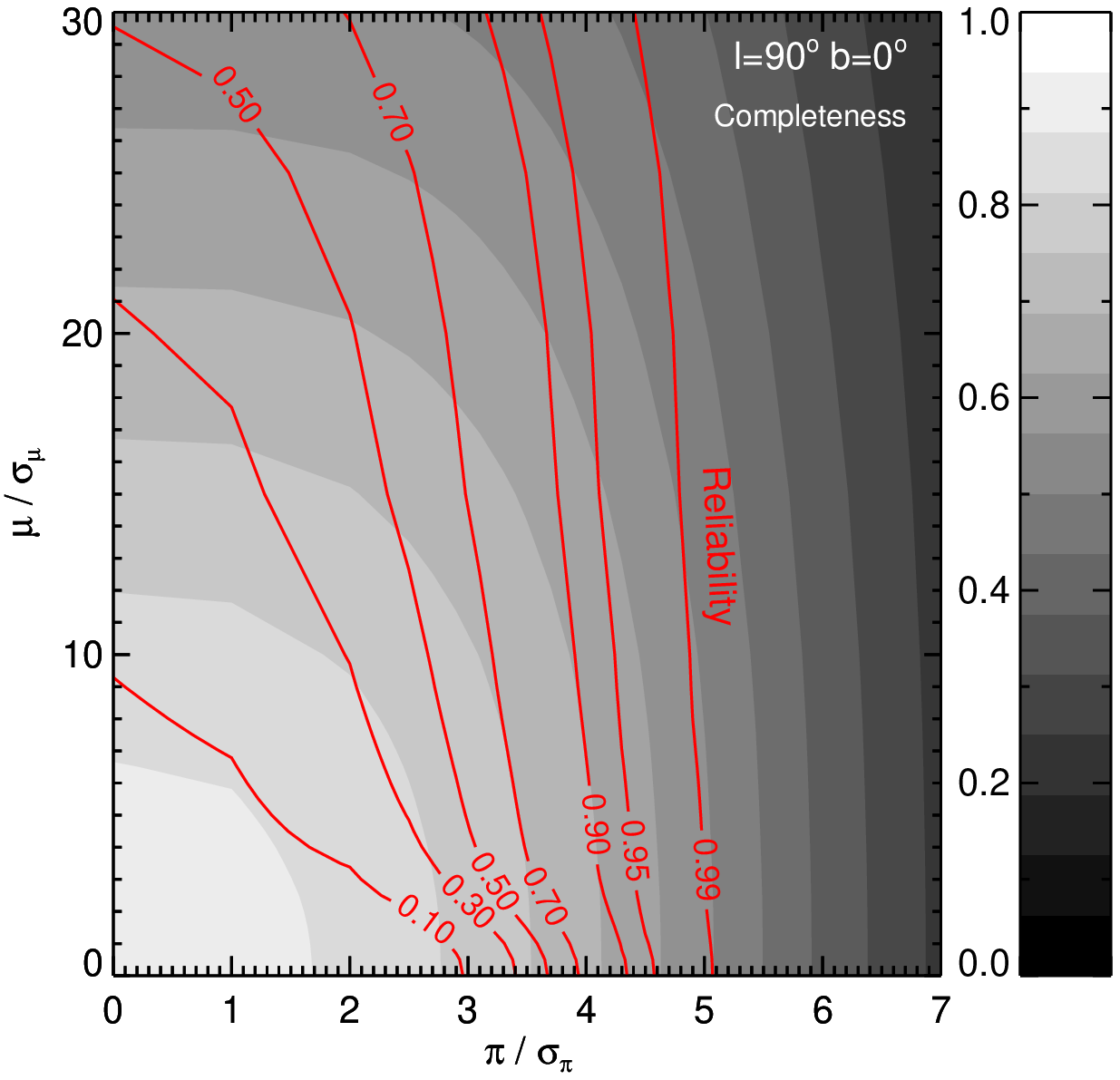}
\includegraphics[height=2.8in]{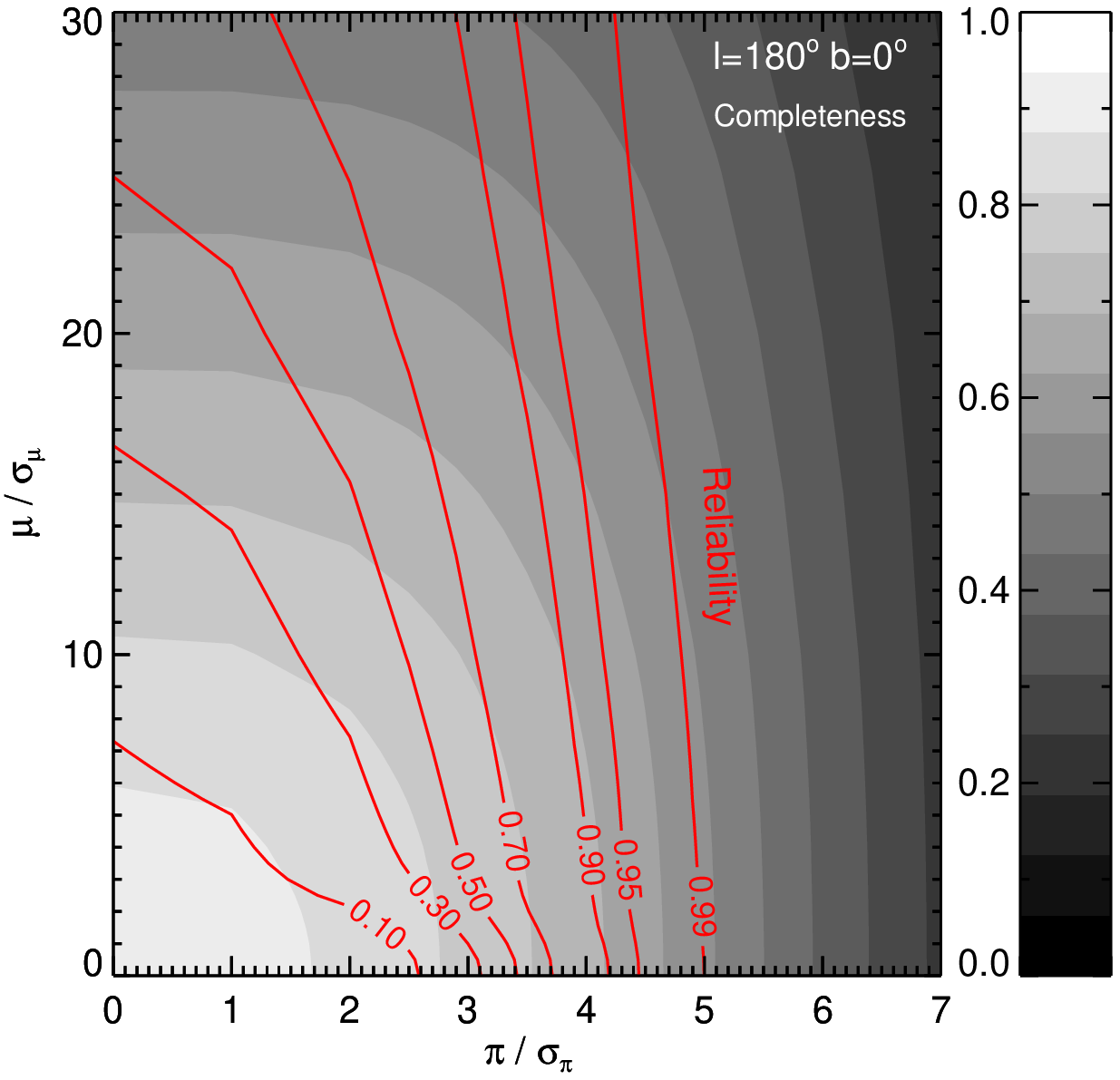}
\includegraphics[height=2.8in]{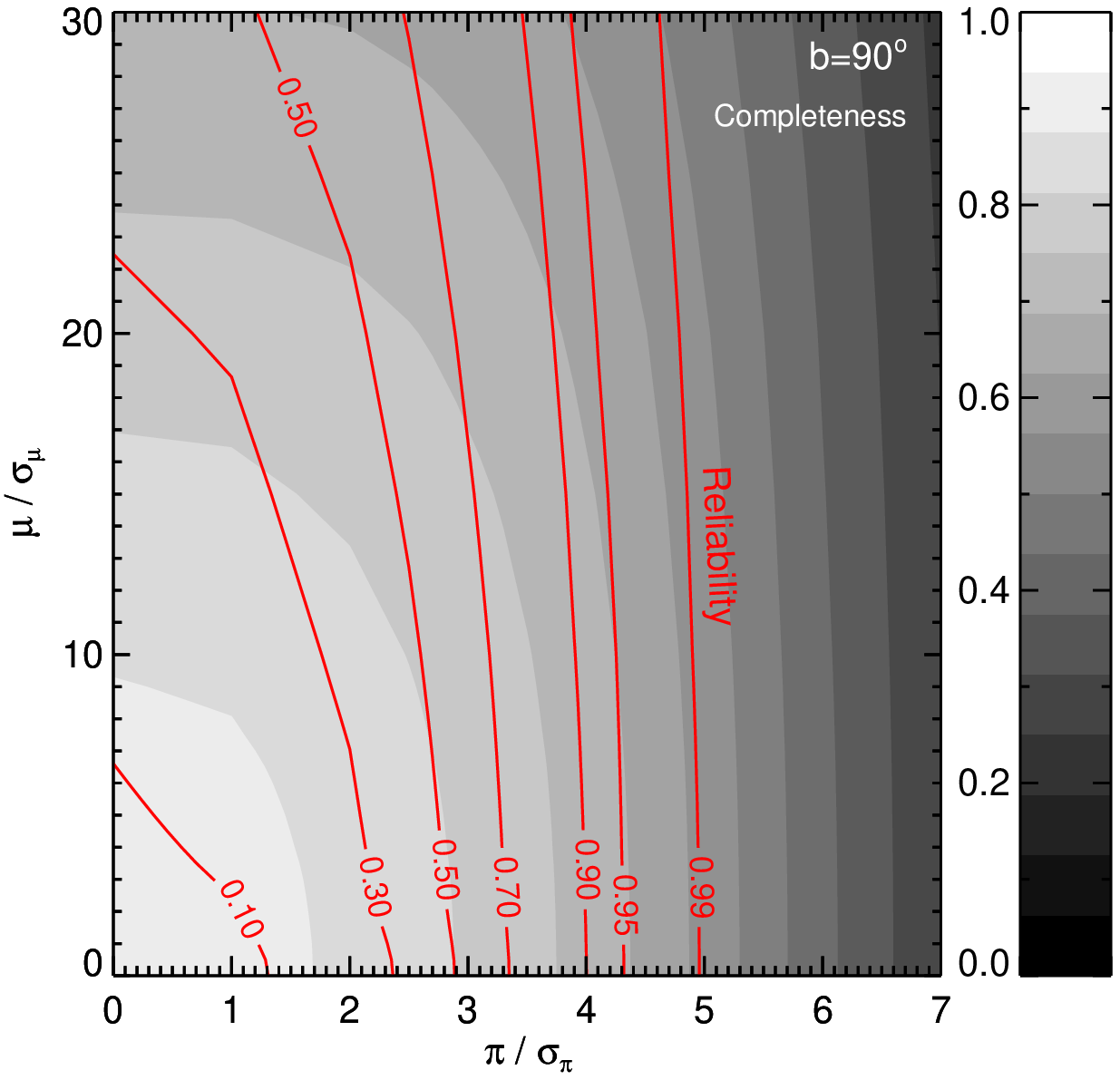}
\caption{The completeness (greyscale) and reliability (red contours) for different selection criteria and lines-of-sight. Proper motion cuts have the greatest benefit for $\pi/ \sigma_\pi > x, ~~\mu / \sigma_\mu > y$ in the range $2 < x < 5$ and  $y< 10$. In this region, more aggressive proper motion cuts improve reliability, and has little impact on completeness.} 
\label{fig:contour}
\end{figure}

Figure \ref{fig:contour} plots the completeness (greyscale) and reliability (red contours) for different selection criteria and fields. Consider first the completeness for each field. As expected, completeness decreases monotonically for increasing cut strength. Note that the $b=90^o$ field is less sensitive to proper motion cuts. This is because proper motion measurements for lines of sight at high latitudes trace motion parallel to the plane, while measurements at low latitudes are more sensitive to vertical motions. Because velocities in the plane are faster than those out of the plane (both in absolute terms and relative to the solar motion), objects towards $b=90^o$ display higher proper motions, and proper motion cuts have less of an effect on completeness in this field. 

The reliability varies more substantially between the three fields. As was seen in Section \ref{sec:lk}, lines of sight closer to the Galactic center display lower reliability for a given selection strategy; this is due to increased contamination from distant objects. Especially in the range $\pi / \sigma_\pi > x,~~ 2 < x < 5$, more aggressive proper motion cuts increase reliability. Furthermore for $\mu / \sigma_\mu >y,~~ y< 10$, these cuts have little impact on completeness. 

\begin{figure}
\includegraphics[width=3.5in]{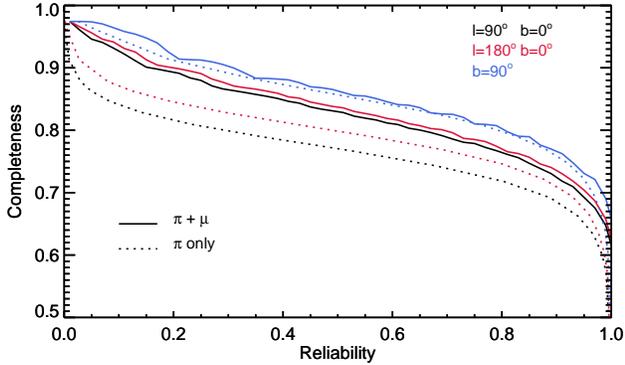}
\caption{The completeness and reliability of optimal strategies. Non-optimal strategies fall below these curves. The solid lines trace strategies using both parallax and proper motion selection, while the dotted lines trace parallax cuts only. The slight jaggedness in the curves is an artifact of our finite sampling of strategies.} 
\label{fig:optimum}
\end{figure}

Figure \ref{fig:contour} presents an opportunity to identify optimal filter strategies, by maximizing completeness along contours of constant reliability (or vice-versa). Figure \ref{fig:optimum} displays the results of this optimization. It shows the maximum completeness obtainable for any given reliability. Also shown in dotted lines are the optimal strategies without proper motion cuts. For most values of reliability, proper motion cuts increase the completeness by $5-10\%$. Because non-optimal strategies fall below these curves -- and because the curves are monotonic -- Figure \ref{fig:optimum} also displays the maximum reliability for a given completeness. Especially for high completeness values, stronger proper motion  cuts improve reliability by \textgreater 10\%. 

Proper motion provides the greatest benefit to lines of sight in the Galactic plane. This is where interloper confusion is most severe, and where additional constraints help to resolve the confusion. Conversely, proper motion filtering has little impact on lines-of-sight out of the plane. 

Table \ref{tab:optimum} lists the specific strategies that lead to the curves in Figure \ref{fig:optimum}. In general, high reliability strategies combine $5 \sigma$ parallax cuts with 1--8 $\sigma$ proper motion cuts, depending on Galactic latitude.

\begin{deluxetable}{ccll}
\tablecolumns{4}
\tablewidth{0in}
\tabletypesize{\scriptsize}
\tablecaption{Optimum Filter Strategies \label{tab:optimum}}
\tablehead{\colhead{Reliability} & \colhead{Max Completeness} & \colhead{$\pi/\sigma_\pi$} &
\colhead{$\mu/\sigma_\mu$}}
\startdata

\multicolumn{4}{c}{$\ell = 90^o ~~~b=0^o$}\\
\hline
0.50 & 0.83 & 2.9 & 6.2 \\
0.60 & 0.81 & 3.2 & 5.5 \\
0.70 & 0.79 & 3.4 & 5.5 \\
0.80 & 0.76 & 3.7 & 5.5 \\
0.90 & 0.72 & 4.1 & 4.5 \\
0.95 & 0.68 & 4.3 & 7.1 \\
0.99 & 0.59 & 4.9 & 8.0 \\

\hline
\multicolumn{4}{c}{$\ell = 180^o ~~~b=0^o$}\\
\hline
0.50 & 0.83 & 2.9 & 4.2 \\
0.60 & 0.82 & 3.1 & 4.0 \\
0.70 & 0.79 & 3.4 & 4.0 \\
0.80 & 0.77 & 3.6 & 4.0 \\
0.90 & 0.73 & 4.0 & 4.0 \\
0.95 & 0.69 & 4.3 & 4.4 \\
0.99 & 0.59 & 4.9 & 5.4 \\

\hline
\multicolumn{4}{c}{$b=90^o$}\\
\hline

0.50 & 0.86 & 2.8 & 3.5 \\
0.60 & 0.84 & 3.0 & 3.0 \\
0.70 & 0.82 & 3.3 & 3.0 \\
0.80 & 0.80 & 3.6 & 2.5 \\
0.90 & 0.76 & 4.0 & 2.0 \\
0.95 & 0.72 & 4.3 & 2.0 \\
0.99 & 0.63 & 5.0 & 1.0 \\

\enddata
\end{deluxetable}

\section{Discussion}
Figure \ref{fig:optimum} demonstrates that parallax and proper motion cuts lead to more complete and reliable solar neighborhood catalogs than strategies based on parallax cuts alone. This is true mainly because proper motion selection is effective at identifying the false parallax signals discussed in Section \ref{sec:lk}. While parallax provides the best quantitative constraint on distance, proper motion is effective at discriminating between distant sources and those in the solar neighborhood.

The main disadvantage of proper motion selection is that, unlike parallax, there is not a one-to-one relationship between proper motion and distance. Because of this, proper motion selection necessarily introduces biases into the extracted catalog. For example, interlopers in a proper-motion selected catalog will be biased towards population II stars, on higher-velocity orbits through the Galactic potential. In addition, nearby sources on orbits similar to the sun's will be culled.

An alternative strategy would incorporate color cuts to suppress interloper confusion. This would affect the resulting catalog in a different way; color cuts would bias against unusually red or blue sources. Depending on the science application, this bias might be less desirable than the bias introduced by proper motion cuts; extreme colors imply unusual and potentially interesting physical conditions in the (sub)stellar atmosphere. 

Finally, we note that while the above simulations illustrate how to design complete and reliable source catalogs, the quantitative performance of each strategy depends on the specific error properties of the data. We have assumed sensible noise parameters describing the \panstarrss 3$\pi$ Survey, but these parameters are likely to evolve as the survey matures. In addition, different scientific motivations will lead to different operational definitions for completeness and reliability. Both of these factors will change the detailed performance of each strategy. Thus, we recommend repeating simulations like these when deriving the final solar neighborhood catalogs from \panstarrss data.

\section{Conclusion}
In this report we have investigated to what extent the \panstarrss $3\pi$ Survey will be able to reliably identify nearby objects (D \textless 100\, pc). We have argued that identifying such objects by parallax alone is not optimal; such a strategy is substantially contaminated by interlopers, due to the fact that nearby sources are so heavily outnumbered by distant ones. Because of these interlopers, \panstarrss parallax measurements will be ineffective at identifying nearby objects more distant than 80 pc (assuming $\sigma_\pi = 2$\,mas; Section \ref{sec:lk}). 

Additional proper motion cuts are able to further discriminate between nearby and distant objects. We have derived a set of optimal strategies incorporating both parallax and proper motion cuts, which maximize completeness for a given level of reliability (and vice-versa). We estimate that 99\% reliable catalogs will be 60\% complete, while catalogs at 50\% reliability will be 85\% complete.

Compared to other possible cuts (e.g. selection on color) we suspect that proper motion cuts are less susceptible to biasing against finding as-yet unidentified classes of objects in the solar neighborhood; these putative objects may well have unusual colors, but would mostly display typical proper motions.

We thank T. Dupuy for helpful discussions about \panstarrss sensitivity calculations, and A. Robin for discussions about the \besancon model. 

\appendix{\section{Synthesizing Pan-STARRS measurements from the Besan{\c c}on catalogs}
The \besancon model generates realistic synthesized catalogs of sources one might observe at a given location on the sky. Because we are interested in the specific performance of the PS1 $3 \pi$ Survey, we determine apparent magnitudes in the \panstarrss (g,r,i,z,y) system and, from these quantities, the corresponding astrometric errors. Relevant parameters describing the \panstarrss PS1 telescope, Gigapixel Camera, and $3\pi$ Survey are provided in Tables \ref{tab:panstarrs} and \ref{tab:panstarrs-filter} \citep{Onaka08, Magnier08, Dupuy09}.

The \besancon catalogs return magnitudes in the CFHT Megacam photometric system. The \panstarrss and Megacam g filters are similar, so we accept the g magnitude included in the catalog by default. To determine fluxes in the other bands, we derive colors for each star by convolving a representative stellar spectrum with the \panstarrss filter throughput profiles. We follow the same process discussed in \citepy{Dupuy09}, and the spectra we use come from \citepy{Pickles98}. We approximate the white dwarfs in the \besancon catalogs as blackbodies, and AGB stars as class I M5 giants. We manually model interstellar reddening via an extinction rate of 0.7 mag $A_V$ kpc$^{-1}$, and an extinction curve from Cardelli et al. (\citeyear{Cardelli89}). Lines of sight are only reddened while they are within 300 pc of the Galactic plane, since this is where most Galactic dust is confined. Note that the photometric quantities themselves are not directly used in our analysis; they are used merely to derive astrometry errors, and to determine detection and saturation limits.

To determine the Poisson noise associated with each source, we convert apparent magnitudes to photo-electron counts given the parameters in Tables \ref{tab:panstarrs} and \ref{tab:panstarrs-filter}. We assume photometry is performed via PSF fitting, such that all photo-electrons from the source are used when estimating the Poisson source noise.

\begin{deluxetable}{rrrrrr}
\tablecolumns{6}
\tablewidth{0in}
\tabletypesize{\scriptsize}
\tablecaption{\panstarrss System Parameters \label{tab:panstarrs}\tablenotemark{a}}
\tablehead{
\colhead{Area} &
\colhead{$\Omega_{\rm fwhm}$} & 
\colhead{$\Omega_{\rm pixel}$} &
\colhead{$\sigma_{\rm read}$} &
\colhead{N$_{\rm epoch}$} & 
\colhead{t$_{\rm baseline}$} \\
\colhead{(m$^2$)} & \colhead{arcsec$^2$} & \colhead{arcsec$^2$} &
\colhead{electrons px$^{-1}$} & \colhead{} & \colhead{years}
}
\startdata
1.73 & 0.8 & 0.26 & 5 & 12\tablenotemark{b, c} & 3.5\tablenotemark{c} \\
\enddata
\tablenotetext{a}{Data from \citealt{Magnier08} and \citealt{Onaka08}.}
\tablenotetext{b}{Epochs per filter. Due to gaps in the focal plane array, the average object will be detected in 80\% of the epochs.}
\tablenotetext{c}{Specific to the $3\pi$ Survey}
\end{deluxetable}

\begin{deluxetable}{llrr}
\tablecolumns{4}
\tablewidth{0in}
\tabletypesize{\scriptsize}
\tablecaption{\panstarrss Filter-Dependent Parameters\tablenotemark{a} \label{tab:panstarrs-filter}}
\tablehead{
\colhead{Band} &
\colhead{t$_{\rm exp}\tablenotemark{b}$} &
\colhead{$\Phi_0$\tablenotemark{c}} &
\colhead{$\Phi_{\rm sky}$\tablenotemark{d}} \\
\colhead{} & \colhead{(s)} & \colhead{($10^9$ s$^{-1}$)} & \colhead{(s$^{-1}$)}}
\startdata
g & 60 & 8.18 & 17.3 \\
r & 38 & 10.2 & 56.5 \\
i & 30 & 9.60 & 122 \\
z & 30 & 6.54 & 195 \\ 
y & 30 & 3.20 & 225 \\
\enddata
\tablenotetext{a}{Data from \citep{Dupuy09}}
\tablenotetext{b}{For the $3\pi$ Survey}
\tablenotetext{c}{Photoelectron rate from a zero magnitude source.}
\tablenotetext{d}{Photoelectron rate for the sky. Assumes that each source measurement samples an effective angular area of $4 \pi \sigma_{\rm psf}^2$, appropriate for PSF fitting \citep{Mighell03}.}
\end{deluxetable}

We include two additional sources of noise: Poisson noise from the sky, and read noise. For PSF-fitting photometry, the effective area of sky sampled for noise estimates is $4\pi \sigma_{\rm psf}^2$ \citep{Mighell03}. The total number of sky photoelectrons measured with each source is given in Table \ref{tab:panstarrs-filter}. By adding the source-, sky-, and read-noise terms in quadrature, we arrive at the photometric signal-to-noise ratio for each measurement.

We model two sources of astrometric error. The first is due to the broadened point spread function (PSF) induced by the atmosphere, which we take to be Gaussian. The ability to determine the center of this Gaussian is given approximately by \citep{Mighell03}
\begin{equation}
\sigma_1 = \frac{\sigma_{\rm psf}}{\rm SNR}
\end{equation}
where $\sigma_{\rm psf}$ is the width of the PSF and SNR is the signal-to-noise ratio derived above. 

The second astrometric noise source reflects the fact that the PSF centers are randomly displaced from their true centers because of turbulent atmospheric refraction \citep{Christou03}. Most of this refraction is correlated among nearby stars, and can be corrected for during data reduction. However, the differential refraction between nearby sources (known as tilt isoplanatism) is not correctable. This adds a random offset to each stellar position, and sets an error floor ($\sigma_2$) for each exposure; this error dominates $\sigma_1$ for bright sources. The \panstarrss error floor is approximately 10 mas per exposure \citep{Magnier08}. Assuming that this is also Gaussian distributed and statistically uncorrelated with $\sigma_1$, the net astrometric error per epoch is given by

\begin{equation}
\label{eq:noise}
\sigma^2 = \frac{\sigma_{\rm psf}^2}{SNR^2} + \sigma_2^2 
\end{equation}

Equation \ref{eq:noise} gives the positional error for a single observation, in a single filter. We are mainly interested in the parallax and proper motion uncertainty, given several position measurements with uncertainty $\sigma$. From the Central Limit Theorem, for sufficiently large $N$ we expect
\begin{equation}
\sigma_\pi \propto \frac{\sigma}{\sqrt{N}}~~
\sigma_\mu \propto \frac{\sigma}{t_{\rm baseline} \sqrt{N}}
\label{eq:stacknoise}
\end{equation}
where $N$ is the number of observations, $\sigma$ is given in equation \ref{eq:noise}, and $t_{\rm baseline}$ is the time baseline of the observations in years (since proper motion is reported in units of angular displacement per year). As Table \ref{tab:panstarrs} indicates, each field will be observed 12 times in each of 5 filters, over a period of 3.5 years. Because gaps fill 20\% of the focal plane array, the average source within this field will be observed 9.6 times. 

We determine the constants of proportionality for Equation \ref{eq:stacknoise} in a Monte Carlo fashion, by fitting parallax and proper motion to many collections of synthetic observations. While $\sigma_\pi$ varies somewhat with position on the sky, the two constants of proportionality are given approximately by 1.3 for $\sigma_\pi$ and 3.5 for $\sigma_\mu$. The proper motion error corresponds to the uncertainty along each direction.

To summarize, then, equation A3, combined with the constants of proportionality given above, give the per-filter, stacked error in parallax and proper motion. These errors can be further reduced by stacking the signal from each filter. We thus compute the standard error of the weighted mean of the five per-filter parallax estimates. This number finally gives the parallax and proper motion uncertainty for each object in the catalog.

As an example, consider a bright source ($\frac{\sigma_{\rm psf}}{SNR} \ll 1$ for every filter). Then, from equation \ref{eq:noise}, $\sigma=\sigma_2=10$\,mas, per filter, per exposure. Using equation \ref{eq:stacknoise}, the net per-filter parallax uncertainty is $\sigma_\pi = 1.3 \times 10 / \sqrt{9.6} = 4.2$\,mas. Since each filter has the same uncertainty in this case, the standard error on the weighted mean of the five per-filter measurements is $\sigma_\pi = 4.2 / \sqrt{5} = 1.88$\,mas. Note that, in general, the five per-filter uncertainties need not be the same -- SNR often varies across filters.

\bibliographystyle{apj}
\end{document}